\documentclass[12pt]{article}

\catcode`\@=11
\@addtoreset{equation}{section}

\global\arraycolsep=2pt
\usepackage{amsbsy,amssymb,latexsym,amsfonts,amsmath}
\usepackage{graphicx,color}

\allowdisplaybreaks

\begin{document}
\begin{flushright}
\parbox{4.2cm}
{UCB-PTH-10/03, IPMU10-0021}
\end{flushright}

\vspace*{0.7cm}

\begin{center}
{\Large \bf 
Universal time-dependent deformations of Schr\"odinger geometry}
\vspace*{2.0cm}\\
{Yu Nakayama}
\end{center}
\vspace*{-0.2cm}
\begin{center}
{\it Berkeley Center for Theoretical Physics, \\ 
University of California, Berkeley, CA 94720, USA \\

and  \\

Institute for the Physics and Mathematics of the Universe, \\
University of Tokyo, Kashiwa, Chiba 277-8582, Japan
}
\vspace{3.8cm}
\end{center}

\begin{abstract} 
We investigate universal time-dependent exact deformations of Schr\"odinger geometry. We present 1) scale invariant but non-conformal deformation, 2) non-conformal but scale invariant deformation, and 3) both scale and conformal invariant deformation. All these solutions are universal in the sense that we could embed them in any supergravity constructions of the Schr\"odinger invariant geometry. We give a field theory interpretation of our time-dependent solutions. In particular, we argue that any time-dependent chemical potential can be treated exactly in our gravity dual approach.

\end{abstract}

\thispagestyle{empty} 

\setcounter{page}{0}

\newpage

\section{Introduction} 
The advent of the AdS/CMP correspondence\footnote{CMP stands for Condensed Matter Physics. See \cite{Hartnoll:2009sz}\cite{McGreevy:2009xe} for recent reviews on the subject.} radically changes the status of the string theory, or quantum gravity. The holography is believed to be one of the fundamental principles of quantum gravity, but so far it has been formidable to acquire any experimental evidence. First of all, our observable universe is unique, so the holographic approach to our universe (if possible) is restricted to one and the only one particular example. On the other hand, we have investigated the string/gravity dual for the QCD, but again our QCD is unique, so experimental comparison of our holographic theory has been quite limited.

The AdS/CMP correspondence has completely changed the situation. We can engineer the condensed matter system as we like, and as a consequence we may have infinitely many experimentally testable holographic setups in principle. We believe that  we will be able to compare various multiverse with condensed matter systems in the near future: we will soon realize that we are surrounded by quantum universes realized in condensed matter systems.

We are, however, still on the way to the above-mentioned paradise of holographic quantum gravities. Unlike the AdS/CFT correspondence and partially successful AdS/QCD correspondence, we do not have any concrete (experimentally testable) realizations of condensed matter systems in terms of quantum gravity. In this sense, we have not reached even the standard of AdS/QCD correspondence, where we can at least compare qualitative predictions with experiments, assuming $N=3$ is large enough. 

In particular, less is known for the field theory dual of the non-relativistic AdS/CFT correspondence. The geometry that has the isometry corresponding to the Schr\"odinger group was first advocated in \cite{Son:2008ye}\cite{Balasubramanian:2008dm},\footnote{A geometric realization of the Schr\"odinger group was pioneered in the earlier work \cite{Duval:1990hj}, whose relation to \cite{Son:2008ye}\cite{Balasubramanian:2008dm} was discussed in \cite{Duval:2008jg}.} and its supergravity embedding has been discussed in the context of the string theory \cite{Herzog:2008wg}\cite{Maldacena:2008wh}\cite{Adams:2008wt}\cite{Hartnoll:2008rs}\cite{Bobev:2009mw}\cite{Donos:2009xc}\cite{Donos:2009zf} as well as in the M-theory \cite{Donos:2009xc}\cite{Ooguri:2009cv}\cite{Jeong:2009aa}. Unfortunately, we do not know the corresponding gauge theories except in some specific cases where the theory is supposed to be obtained from the discrete light cone quantization (DLCQ) of the $\mathcal{N}=4$ super Yang-Mills theory. The DLCQ is notoriously difficult to study, so in practice, we do not have any calculable Lagrangian description of the dual field theory for the non-relativistic AdS/CFT correspondence.

In this paper, as a first step to understand the nature of the non-relativistic AdS/CFT correspondence, we study the universal time-dependent deformations of the simplest Schr\"odinger invariant geometry. Our solutions are exact and universal in the sense that we can embed them in any supergravity constructions of Schr\"odinger invariant geometry. We study the field theory interpretations of our deformations, and we claim that the field theories dual to the Schr\"odinger invariant geometry should always admit such time-dependent exact deformations of the action, and they should also possess the states corresponding to our time-dependent solutions. The analysis furthermore reveals that the dual field theories always include certain operators that are not contained in the minimal operator contents of the Schr\"odinger invariant field theory. In other words, the Schr\"odinger invariant field theories that have a gravity dual predict an existence of these particular operators. 

As a spin-off of our results, we argue that any time-dependent chemical potential can be treated exactly in our gravity dual approach. The time-dependent chemical potential turns out to be simply the time-dependent coordinate transformation of the bulk theory. 

Our exact time-dependent deformations will show PP-wave singularities from the bulk gravitational theory viewpoint. It would be interesting to understand the nature of the singularity from the field theory perspective. Since we have not succeeded in finding the complete dual field theories, we cannot say much about the fate of the singularity from the field theory viewpoint. We hope that once the  non-relativistic AdS/CFT correspondence is much better established, we would be able to attach the resolutions of PP-wave singularities from the dual non-relativistic field theory viewpoint. We leave this important issues for the future study.

\section{Universal Deformations of Schr\"odinger space-time}

We begin with the geometry with the Schr\"odinger invariance \cite{Son:2008ye}\cite{Balasubramanian:2008dm}:
\begin{align}
ds_{d+3}^2 = -2 \frac{dt^2}{z^4} + \frac{-2dt d\xi + dx_i^2 + dz^2}{z^2} \ , \label{schr}
\end{align}
where $i=1,\cdots d$. The light-like $\xi$ direction is compactified as $\xi \simeq  \xi + 2\pi R$ so that the spectrum reproduces the quantization of the particle number in the dual field theory.
The metric \eqref{schr} is the solution of the Einstein equation with the massive (Proca) vector field
\begin{align}
S = \int d^d x_i d\xi dt dz \sqrt{-g} \left(\frac{1}{2}R - \Lambda - \frac{1}{4} F_{\mu\nu} F^{\mu\nu} - \frac{m^2}{2}A_\mu A^\mu \right) \ . \label{action}
\end{align}
where $F_{\mu\nu} = \partial_\mu A_\nu - \partial_\nu A_\mu$. One can show that  $A = A_\mu dx^\mu = -\frac{dt}{z^2}$ solves the equation of motion as well as the Einstein equation, provided
\begin{align}
\Lambda = -\frac{1}{2}(d+1)(d+2) \ , \ \ m^2 = 2(d+2) \ .
\end{align}

The geometry has an obvious invariance under the translation in $(t,x^i)$ as well as the Euclidean rotation in $x^i$. It is also invariant under the Galilean boost
\begin{align} x^i \to x^i - v^it \ , \ \ \xi \to \xi - v^i x^i + \frac{1}{2} v^2 t \ .
\end{align}
 Furthermore, the geometry has the full non-relativistic conformal invariance \cite{Hagen:1972pd}\cite{Niederer:1972zz}.
The dilatation is generated by 
\begin{align}
t \to \lambda^2 t \ , \ \ x^i \to \lambda x^i \ , \ \ z \to \lambda z\ , \ \ \xi \to \xi
. \label{dilatation}
\end{align}
 The non-relativistic special conformal transformation is generated by 
\begin{align}
t\to \frac{t}{1+at}\ , \ \ x^i \to \frac{x^i}{1+at}\ , \ \  z\to \frac{z}{1+at} \ , \ \  \xi \to \xi -\frac{a}{2}\frac{x^ix^i +z^2}{1+at} \ . \label{conformal}
\end{align}
\cite{SchaferNameki:2009xr} argues that the metric \eqref{schr} is the simplest geometrical realization of the Schr\"odinger invariance within the coset space construction. 

The geometry is proposed to be dual to a non-relativistic Schr\"odinger invariant field theory. So far, there is no concrete proposal for what is the precise dual field theory corresponding to the geometry. It is suggested in \cite{Son:2008ye} that it would be obtained by a relevant deformation of the relativistic CFT dual to the $AdS_5$ space:\footnote{Note that the time $t$ here is a light-cone time $x^+$ in the original relativistic AdS/CFT coordinate.}
\begin{align}
 \delta S_{\text {CFT}} =  J\int d^{d+2}x O^t  \label{dGKP}
\end{align}
after compactifying a light-cone direction because the GKPW prescription dictates that the deformation \eqref{dGKP} is induced by the Proca field $A_\mu$. The relativistic scaling dimension of $O^\mu$ is $d+2$, where $O^\mu$ is dual to $A_\mu$. 

One of the main objectives of this paper is to understand the physics of this Schr\"odinger invariant geometry by introducing the exact time-dependent deformations. In section 2.1, we first study the scale invariant but non-conformal deformations, and in section 2.2, we study the conformal but non-scale invariant deformations. The both deformations sound peculiar from our experience in relativistic field theories: indeed, we would like to argue that they are special features of time-dependent non-relativistic field theories. In section 2.3, we extend our analysis to broader classes of exact time-dependent solutions.

\subsection{Scale invariant but non-conformal deformations}
We begin with the scale invariant but non-conformal deformations of the Schr\"odinger geometry. It was shown \cite{Nakayama:2009ww} that this is impossible without breaking further symmetries, so we investigate the solutions that are time-dependent explicitly.
The most general scale invariant deformations (up to a coordinate transformation) are given by
\begin{align}
ds_{d+3}^2 =  -C\left(\frac{t}{z^2}\right) \frac{dt^2}{z^4} + D\left(\frac{t}{z^2}\right)\frac{-2dt d\xi + dx_i^2 + dz^2}{z^2}
\end{align}
for the metric and 
\begin{align}
A = A_\mu dx^\mu = -E\left(\frac{t}{z^2}\right)\frac{dt}{z^2} \ ,
\end{align}
for the vector field. It is easy to see that $t/z^2$ is invariant under the scale transformation \eqref{dilatation} but not under the non-relativistic special conformal transformation \eqref{conformal}.
We have used the diffeomorphism invariance to remove the $dzdt$ component of the metric. 

The $(zt)$ component of the Einstein equation tells us $D = \text{const} $, so we set $D=1$. 
The other equations of motion can be solved exactly by
\begin{align}
 C &= 2 + C_1 \frac{z^2}{t} + C_2 \left(\frac{z^2}{t}\right)^{d/2+2} + \epsilon^2\frac{d+2}{d+3} \frac{z^{2d+8}}{t^{d+4}} \cr
E &= 1 + \epsilon \frac{z^{d+4}}{t^{d/2+2}} \ ,
\end{align}
where $C_1$, $C_2$ and $\epsilon$ are integration constants. 

The deformations by $C_1$ and $C_2$ do not depend on the background vector field, and they even exist without introducing the vector field ({\it e.g.} in locally AdS background).  
Such PP-wave deformations in the AdS space was studied in \cite{Chamblin:1999cj}\cite{Brecher:2000pa}\cite{Kumar:2004jv}\cite{Sfetsos:2005bi}.\footnote{See also \cite{Chu:2006pa}\cite{Das:2006dz}\cite{Das:2006pw} for related null deformed backgrounds.} The metric is given by
\begin{align}
ds_{d+3}^2 =  -\left( C_1 \frac{z^2}{t} + C_2 \left(\frac{z^2}{t}\right)^{d/2+2}\right) \frac{dt^2}{z^4} + \frac{-2dt d\xi + dx_i^2 + dz^2}{z^2} \ ,
\end{align}
which solves the vacuum Einstein equation with the negative cosmological constant.
In the AdS case, it was argued in \cite{Brecher:2000pa} that the deformation by $C_2$ is dual to the time-dependent vacuum expectation value (VEV) for a particular component of the energy momentum tensor:\footnote{Again, note that $t$ here corresponds to the light-cone direction $x^+$ in the original relativistic AdS/CFT coordinate.}
\begin{align}
\langle T_{tt} \rangle = \frac{C_2}{t^{d/2 + 2}} \  . 
\end{align}
Later, we will discuss the similar operator has a VEV in the dual Schr\"odinger invariant field theory.

It is sometimes believed that the scale invariance implies the conformal invariance. This is not always true because there is no symmetric reason why this is so \cite{Coleman:1970je} a-priori, and indeed there are some known counterexamples \cite{Hull:1985rc}\cite{Riva:2005gd}\cite{Chu:2006pa}. However, it was shown that in $(1+1)$ dimension, the Poincar\'e invariance, unitarity, and the discreteness of the spectrum guarantees the equivalence between the conformal invariance and the scale invariance \cite{Zamolodchikov:1986gt}\cite{Polchinski:1987dy}. It is hoped that a similar statement should hold in higher dimensions with a suitable generalization of the assumptions \cite{Nakayama:2009qu}\cite{Nakayama:2009fe}\cite{Dorigoni:2009ra}.
In the unitary Sch\"odinger invariant field theories, it is conjectured that the scale invariance together with the Galilean invariance, rotation and translational invariance would imply the full non-relativistic conformal invariance \cite{Nakayama:2009ww}. It is easy to see that there is no such time-independent deformations in the above simple Sch\"odinger invariant geometry.\footnote{In higer dimensional supergravity embeddings, there could exist other terms that are scale invariant but not conformal invariant. We also note that the discussion here specializes in the case with the dynamical critical exponent $\mathcal{Z}=2$. When $\mathcal{Z} \neq 2$, the non-relativistic special conformal transformation cannot be constructed in the algebra. See \cite{Adams:2008wt}\cite{Hartnoll:2008rs}\cite{Singh:2009tq}\cite{Donos:2009xc} for examples of such geometries.} On the other hand, in the discussion above, the time-translational invariance is explicitly broken so that the conjecture does not apply.

\subsection{Conformal but not scale invariant deformations}
Now, we will present a more peculiar situation where the solution is invariant under the non-relativistic conformal transformation but not invariant under the dilatation. Such a geometry is impossible without breaking a further symmetry, and here again, we consider explicitly time-dependent solutions.
The most general non-relativistic conformal invariant deformations of the Schr\"odinger geometry are given by
\begin{align}
ds_{d+3}^2 =  -C\left(\frac{t}{z}\right) \frac{dt^2}{z^4} + D\left(\frac{t}{z}\right)\frac{-2dt d\xi + dx_i^2 + dz^2}{z^2}
\end{align}
for the metric and 
\begin{align}
A = -E\left(\frac{t}{z}\right)\frac{dt}{z^2} \ ,
\end{align}
for the vector field. It is easy to see that $t/z$ is invariant under the non-relativistic special conformal transformation \eqref{conformal} but not under the scale transformation \eqref{dilatation}. Again, we have used the diffeomorphism invariance to remove the $dzdt$ component of the metric. 

The $(zt)$ component of the Einstein equation tells us $D = \text{const} $, so we set $D=1$. 
The other equations of motion can be solved exactly by
\begin{align}
 C &= 2 + C_1 \frac{z^2}{t^2} + C_2 \left(\frac{z}{t}\right)^{d+4} + \epsilon^2\frac{d+2}{d+3} \frac{z^{2d+8}}{t^{2d+8}} \cr
E &= 1 + \epsilon \frac{z^{d+4}}{t^{d+4}} \ ,
\end{align}
where $C_1$, $C_2$ and $\epsilon$ are integration constants. The deformations by $C_1$ and $C_2$ do not depend on the background vector field, and they even exist without introducing the vector field ({\it e.g.} in locally AdS background).  

Such PP-wave deformations in the AdS space was studied in \cite{Chamblin:1999cj}\cite{Brecher:2000pa}\cite{Kumar:2004jv}\cite{Sfetsos:2005bi}. The metric is given by
\begin{align}
ds_{d+3}^2 =  -\left( C_1 \frac{z^2}{t^2} + C_2 \left(\frac{z}{t}\right)^{d+4}\right) \frac{dt^2}{z^4} + \frac{-2dt d\xi + dx_i^2 + dz^2}{z^2} \ .
\end{align}
It was argued \cite{Brecher:2000pa} that the deformation by $C_2$ is due to the time-dependent VEV for a particular component of the energy momentum tensor:
\begin{align}
\langle T_{tt} \rangle = \frac{C_2}{t^{d + 4}} \  . 
\end{align}

The closure of the non-relativistic conformal algebra ({\it i.e.} $i[H,K] = D$: see appendix A for details) demands that the non-relativistic special conformal invariance implies the scale invariance. Thus, it is impossible to break the  scale invariance without spoiling the special conformal invariance. Here, the explicit time-dependence alleviates the situation, and without the conserved Hamiltonian, the conformal invariance without the dilation is possible.

\subsection{More general deformations}
Although the symmetry is more reduced, our deformations in the previous subsections are particular cases of more general time-dependent exact solutions of the simplest action \eqref{action} that allows the Schr\"odinger invariant geometry.
The more general time-dependent solutions are given by
\begin{align}
ds_{d+3}^2 =  -C(t,z) \frac{dt^2}{z^4} + D(t,z)\frac{-2dt d\xi + dx_i^2 + dz^2}{z^2}
\end{align}
for the metric and 
\begin{align}
A = -E(t,z)\frac{dt}{z^2} \ ,
\end{align}
where 
\begin{align}
C(t,z) &= 2 + z^2 c_1(t) + c_2(t) z^{d+4} + \epsilon(t)^2 \frac{d+2}{d+3} z^{2(d+4)} \cr
D(t,z) &= 1 \cr
E(t,z) &= 1 + \epsilon(t) z^{d+4} \ .
\end{align}
$c_1(t)$, $c_2(t)$ and $\epsilon(t)$ are {\it arbitrary} functions of $t$.

There are several special choices of these functions:
\begin{itemize}
\item scale invariance: $c_1(t) = \frac{1}{t}$, $c_2(t) = \frac{1}{t^{\frac{d+4}{2}}}$ , $\epsilon(t) = \frac{1}{t^{d+4}}$ \ .
\item conformal invariance:  $c_1(t) = \frac{1}{t^2}$, $c_2(t) = \frac{1}{t^{{d+4}}}$ , $\epsilon(t) = \frac{1}{t^{2(d+4)}}$ 
\item scale and conformal invariance: $c_1(t) = \delta(t)$. 
\end{itemize}
We have discussed the first two choices in the previous subsections. The last one may need a comment. The solution contains a delta function and it looks singular, but in our approach, this is the only possible solution that is scale invariant and conformal invariant at the same time. 
As we would like to discuss it later, however, the deformation by $c_1(t)$ can be gauged away by a coordinate transformation, so in practice, there is no non-trivial scale and conformal invariant time-dependent deformation of the Schr\"odinger invariant geometry.

The impossibility to obtain a scale and conformal invariant time-dependent solution may be regarded as a dual statement of the impossibility to construct a scale invariant but non-conformal field theory without breaking any translation invariance. The point is that the non-relativistic conformal algebra has a non-trivial involution anti-automorphism, which exchanges the Hamiltonain $H$ and the non-relativistic special conformal transformation $K$ (see appendix A for details). Therefore, from the representation theory viewpoint, it is very close to study  the theory with no conserved Hamiltonian but invariant under all the other generators including the conformal transformation and to study the theory with no conformal invariance but invariant under all the other generators including the Hamiltonian.


Let us briefly discuss the singularity structure of the geometry. First of all, it is not difficult to see that all the curvature invariants are constant irrespective of the deformations. However, there is a PP-wave singularity at $z=0$ as well as $t=0$ (or $t = \pm \infty$) when $c_2(t)$ or $\epsilon(t)$ becomes infinite as $t\to 0$ (or $t = \pm \infty$).\footnote{The apparent singularity due to $c_1(t)$ can be removed by the coordinate transformation. We will give a field theory interpretation of this removal of singularity later in section 3.} In particular, the latter condition applies at $t=0$ to the scale/conformal invariant deformations discussed in section 2.2 and 2.3. 

Since our deformations vanish as $z\to 0$, the singularity at $z=0$ is the same as that of the original Sch\"odinger space-time. See for example \cite{Hartnoll:2008rs} for the analysis of the singularity at $z=0$. On the other hand, the analysis of the PP-wave singularity structure has been done in the vacuum AdS PP-wave geometry in \cite{Chamblin:1999cj}. Our metric is within the analysis done in \cite{Chamblin:1999cj} except that our solution does not satisfy the vacuum Einstein equation but rather it has the source term from the Proca field. The PP-wave singularity appears whenever
\begin{align}
\mathcal{A} = z^5\partial_z\left(\frac{\partial_z(z^2C(t,z))}{z}\right)
\end{align}
is diverging, and this is precisely the above condition ({\it i.e.} diverging $c_2(t)$, $\epsilon(t)$ at $t=0$). Again note that the deformation by $c_1(t)$ does not introduce any PP-wave singularity at all.

\section{Field theory interpretation}
In this section, we will study the field theory interpretation of the exact time-dependent deformations introduced in the last section. We use the natural generalization of the GKPW prescription \cite{Gubser:1998bc}\cite{Witten:1998qj} of the AdS/CFT correspondence to our non-relativistic setup. See \cite{McGreevy:2009xe}\cite{Son:2008ye}\cite{Fuertes:2009ex}\cite{Volovich:2009yh}\cite{Leigh:2009ck} for its successful applications in non-relativistic AdS/CFT correspondence.

The deformation by $c_1(t)$ corresponds to the introduction of the time-dependent chemical potential in the action:
\begin{align}
M^2\int dt d^{d}x c_1(t) \rho(t,x) \ , \label{chemi}
\end{align}
where $\rho(t,x)$ is the particle number density. For instance, in the free Schr\"odinger theory with the action
\begin{align}
S = \int dt d^d x \left(i\Phi^* \partial_t \Phi - \frac{1}{2M}\partial_i \Phi^* \partial_i \Phi \right) \ ,
\end{align}
the particle number density is given by $\rho = \Phi^\dagger \Phi$. This identification was first proposed in \cite{Son:2008ye}, and we will confirm the identification by comparing the field theory expectation and the gravity prediction. In the local Schr\"odinger invariant field theories, such an operator always exists because it generates the particle number operator $N = \int d^d x \rho$ that lies in the non-relativistic conformal algebra (see Appendix A).

In the free Schr\"odinger theory, the introduction of this deformation modifies the two-point function as
\begin{align}
\left\langle \Phi(t_1,x_1) \Phi^\dagger(t_2,x_2) \right\rangle = \frac{1}{(t_1-t_2)^{d/4}} \exp\left(iM\frac{(x_1-x_2)^2}{t_1-t_2} + i M^2\int_{t_2}^{t_1} dt c_1(t) \right) \  \label{tptf}
\end{align}
up to a proportional factor.
It is important to notice that in the simple Schr\"odinger invariant field theories with Lagrangian description, the time-dependent chemical potential \eqref{chemi} can be exactly treated. The introduction of \eqref{chemi} is equivalent to the field redefinition $\Phi(t,x_i) \to e^{iM^2\int^t ds c_1(s)} \Phi(t,x_i)$. We will argue this is also true whenever the theories have a gravity dual.

To probe the implication of the gravity deformation by $c_1(t)$, let us consider a minimally coupled scalar field $\phi$ with the gravity: 
\begin{align}
S_{\phi} = \int dt d\xi dzd^{d} x \sqrt{-g}\left( \partial^\mu \phi^* \partial_\mu \phi - m_0^2 |\phi|^2 \right) \ .
\end{align}
The equation of motion for $\phi$ is given by 
\begin{align}
\partial_z^2 \phi - \frac{1}{z}(d+2) \partial_z \phi + \left(2iM\partial_t - k^2 - M^2c(t) - \frac{m^2}{z^2} \right) \phi = 0 ,
\end{align}
where $M$ is the momentum eigenvalue in $\xi$ direction, and $k_i$ is that for the $x^i$ directions: $\phi(z,t,x,\xi) = \phi(z,t) e^{ikx+iM \xi}$. $m^2 = m_0^2 + 4M^2$ is the effective mass of the KK mode in the light-cone direction $\xi$. 
We can use the technique of separation of variables to solve the scalar equation of motion. It turns out that the superpositions of the wavefunction:
\begin{align}
\phi = z^{d/2 +1} K_\nu(pz) e^{-iEt+iM^2\int^t ds c_1(s)} \ ,
\end{align}
where $p = \sqrt{k^2-2ME}$, will give a complete basis of the solution. 
In practice, this simply modifies the time-dependence of the scalar field by the phase factor $e^{iM^2\int^t ds c_1(s)}$ compared with the undeformed $c_1(t)=0$ case. Explicitly, the two-point function can be computed as 
\begin{align}
\left\langle O(t_1,x_1) O^\dagger(t_2,x_2) \right \rangle &= \int dE dp e^{-iE(t_1-t_2) + ip(x_1-x_2)} (2ME-p^2)^\nu e^{iM^2\int_{t_1}^{t_2} dt c_1(t) } \cr
& =  \frac{C_{\nu}}{({t_1-t_2})^{\frac{d+2}{2}+\nu}} \exp\left(iM\frac{(x_1-x_2)^2}{t_1-t_2} + i M^2\int_{t_2}^{t_1} dt c_1(t) \right) \ , 
\end{align}
where $C_\nu$ is a (cut-off dependent) constant.
We note the exact agreement with \eqref{tptf}. 

This simplicity of the correlation functions actually suggests that the deformation by $c_1(t)$ is only apparent. To see it, we note that the coordinate transformation $\xi \to \xi - \frac{1}{2}\int^t ds c_1(s)$ will lead us to the original metric. It means that the time-dependent chemical potential in the Schr\"odinger invariant theory is always solved by the coordinate transformation in the dual gravity.

We emphasize that even though in the simplest field theory examples, the time-dependent chemical potential can be introduced/removed by the time-dependent field redefinition $\Phi(x,t)\to e^{iM^2\int^t ds c_1(s)} \Phi(x,t)$, it is not at all obvious this is always the case, when the system is strongly coupled and the Schr\"odinger symmetry is rather emergent. We have, on the other hand, showed that it is always possible to introduce the time-dependent chemical potential in the gravity approach. This will further constrain the candidates of the field theory duals.

The interpretation of $c_2(t)$ is given by assigning the VEV to a certain operator in the dual field theory:
\begin{align}
\langle T_{tt} \rangle = c_2(t) \ . 
\end{align}
The scaling dimension of $T_{tt}$ is $d+4$. This operator does not always exist in non-relativistic Schr\"odinger invariant field theory, but the geometric construction requires the existence. The reason why we call it $T_{tt}$ is that it is exactly the $tt$ component of the energy-momentum tensor in the DLCQ construction of the Schr\"odinger invariant field theories. In the free Schr\"odinger theory, it is given by $\partial_t\Phi^\dagger \partial_t \Phi$. 

Finally, $\epsilon(t)$ corresponds to the VEV of a certain (non-universal) operator of the non-relativistic CFT in the first oder approximation:\footnote{It looks peculiar that there is no corresponding (dual) operator insertion in the action. That would correspond to just the change of scale for $t$. Equivalently, the universal deformation by $\int dt d^dx \mathcal{H}$ does this job, where $\mathcal{H} = T_{00}$ is the Hamiltonian density. See the discussion in the following. Note that this $T_{00}$ is different from $T_{tt}$ introduced above. In the free Schr\"odinger theory, $T_{00} = \frac{1}{2m}\partial_i\Phi^\dagger\partial_i \Phi$. }
\begin{align}
\langle O_{t} \rangle  = \epsilon(t) \ .
\end{align}
The scaling dimension of $O_t$ is $2(d+4)$.
In the DLCQ of the relativistic conformal field theory, we have discussed how $O^\mu$ was introduced to generate the Schr\"odinger invariant geometry. Here, the same operator has been given the VEV. Later, we will discuss the supergravity embedding of our deformations, and there, we will see the origin of the effective vector field corresponding to $O_t$. Various form fields in supergravity play the role of the operator.

So far, we have discussed the field theory interpretation of the exact time-dependent deformations of the simplest Schr\"odinger invariant field theory. Now, we would like to reverse the logic and ask the question: what would we expect for the universal deformations from the field theory viewpoint?
The minimal ingredient of the Schr\"odinger invariant field theory admits two universal deformations that preserve the Galilean boost and rotational invariance. They are given by the density operator $\rho$ and the trace of the energy momentum tensor $T^i_{\ i}  = 2T_{00}$. The latter, however, only changes the normalization of the kinetic term with respect to the potential terms, so the universal deformations are simply given by adding the chemical potential in agreement with the gravity discussion above.

The other two deformations by $c_2(t)$ and $\epsilon(t)$ are not in the minimal list of the non-relativistic conformal field theory. Thus, the non-relativistic conformal field theory dual to the simplest Schr\"odinger invariant predicts the extra existence of the universal operators. In later section 4, we will discuss how these universal deformations appear in the general supergravity embedding of the Schr\"odinger invariant geometry.

\subsection{more on the correlation functions}
We have studied the exact modifications of correlation functions induced by $c_1(t)$.
The deformations given by $c_2(t)$ and $\epsilon(t)$ are more subtle than the deformation by $c_1(t)$ in terms of the correlation functions. These deformation will only affect the IR behavior of the scalar field in the geometry. To understand the situation, let us again consider the minimally coupled scalar field under the $c_2(t)$ deformation. The same analysis applies for $\epsilon(t)$ deformation. The equation of motion is
\begin{align}
\partial_z^2 \phi - \frac{1}{z}(d+1) \partial_z \phi + \left(2iM\partial_t - k^2 - M^2c_2(t)z^{d+2} - \frac{m^2}{z^2} \right) \phi = 0 . \label{seqq}
\end{align}

The UV behavior ({\it i.e.} $z\to 0$) of the scalar field is the same as in the undeformed case:
\begin{align}
\phi \sim e^{ipx-iwt}z^{d/2+1} \left(\frac{I_{-\nu}(|k|z) +A(k)I_\nu(|k|z)}{I_{-\nu}(|k|\epsilon) +A(k)I_\nu(|k|\epsilon)} \right) \ ,
\end{align}
where $\epsilon$ is the cutoff, $k^2 = p^2- 2M\omega$, and $\nu = \frac{1}{2}\sqrt{(d+2)^2 + 4m^2}$. In the undeformed case with $c_2(t) = 0$, the regularity of the Wick-rotated wavefunction at $z\to \infty$ determines $A(k)$ so that $\phi$ is given by $K_\nu(|k|z)$. Here the equation itself is modified toward $z\to \infty$ and in addition we do not have a good guiding principle to set the boundary condition there in our explicitly time-dependent setup. 

To investigate the behavior $z\to \infty$ and study the boundary condition, we need to solve the equation toward $z\to \infty$.
Unfortunately, there is no analytic solution of the equation \eqref{seqq} except for $M=0$. When $M=0$, the solution is uniquely given by 
\begin{align}
\phi \sim e^{ipx-iwt}z^{d/2+1} \frac{K_\nu(|p|z)}{K_\nu(|p|\epsilon)} \ .
\end{align}
The wavefunction with $M=0$ corresponds to the zero-norm state of the field theory, and the two-point functions between such states are not affected at all by the deformation $c_2(t)$ (and $\epsilon(t)$ by repeating the same argument). See \cite{Nakayama:2009ed} for the discussions on the zero-norm states in the Schr\"odinger invariant field theories and their gravity dual.

Instead of giving the analytic solutions of \eqref{seqq}, we can study the power series solution:
\begin{align}
\phi = z^{d+2-\Delta}\sum_{n=0}^{\infty} a_n(t)e^{ikx} z^n \ + z^{\Delta}\sum_{n=0}^{\infty} b_n(t) e^{ikx} z^n \ ,
\end{align}
where $\Delta$ is related to the conformal dimension of the operator and it is given by $\Delta = \frac{d+2}{2}+ \nu = \frac{d+2+\sqrt{(d+2)^2 + 4m^2}}{2}$. 
The equations of motion will determine the higher powers $a_n(t)$ and $b_n(t)$ from the boundary data $a_0(t)$ and $b_0(t)$. For instance,
\begin{align}
a_2(t) = -\frac{2iM\partial_ta_0(t)-k^2 a_0(t)}{4(\nu-1)} \ , \ \ b_2(t) = -\frac{2iM\partial_t b_0(t)-k^2 b_0(t)}{4(\nu+1)} \ ,
\end{align}
and the higher terms are determined recursively.
The series coincide with the power expansion of the Bessel function, up to the $(d+2)$-th order. At the $(d+4)$-th order, they show a deviation:
\begin{align}
\delta a_{d+4}(t) &= \frac{M^2 c_2(t)}{2(\Delta-d-3)(d+4)} a_0(t) \cr
\delta b_{d+4}(t) &=-\frac{M^2 c_2(t)}{2(\Delta +1)(d+4)} b_0(t) \ .
\end{align}
Similarly, with the $\epsilon(t)^2$ deformation of the metric, the deviation appears at the $2d+8$-th order: 
\begin{align}
\delta a_{2d+8}(t)&= -\frac{M^2 \epsilon(t)^2(d+2)}{2(d+3)(d+4)(3d+10-2\Delta)} a_0(t) \cr
\delta b_{2d+8}(t) &=-\frac{M^2 \epsilon(t)^2(d+2)}{2(d+3)(d+6)(d+4+\Delta)} b_0(t) 
\end{align}
by assuming that the vector field does not affect the equation of motion for the scalar.

The special case of the Kaigorodov space \cite{Kaigorodov1}\cite{Kaigorodov2}, where $c_2(t) = c_2 = \text{const}$ was discussed in \cite{Brecher:2000pa}. It was argued that the two-point function does not show any {\it explicit} dependence on $c_2 $ up to contact terms, so we can {\it choose} the same boundary condition as in the case $c_2(t) = 0$, which will yield the independence of the two-point function on $c_2$. Similarly, when $\epsilon(t)$ is a constant, the deformation only gives {\it explicit} corrections to the contact terms, so we can {\it choose} the same boundary condition as in the case $\epsilon(t) = 0$, and we obtain no dependence on $\epsilon$ in the two-point functions.

Formally, one can study the perturbative corrections to the two-point functions by using the Witten diagram and computing the perturbative correction from the ``interaction" $\int dt dz d^{d}x c_2(t) z M^2|\phi|^2$. We do not know how to compute the same perturbative corrections from the field theory side because the expectation value of $T_{tt}$ alone does not seem to specify the ``state" we evaluate the correlation function. Even in the time-independent Kaigorodov case, we do not know the precise prescription to compute the correlation functions from the field theory side. Further studies are needed in this direction to reveal the nature of $c_2(t)$ and $\epsilon(t)$ deformations.

\section{Supergravity embedding}

\subsection{General argument}
We have discussed the three independent universal deformations ({\it i.e.}  $c_1(t)$, $c_2(t)$ and $\epsilon(t)$) of the simplest Schr\"odinger geometry given by the metric \eqref{schr}, which is supported by a free Proca field with the definite mass. It would be interesting to see whether such deformations can be embedded in the string/M-theory solutions. 

There are several supergravity solutions that possess the Schr\"odinger isometry with \cite{Hartnoll:2008rs}\cite{Bobev:2009mw}\cite{Donos:2009xc}\cite{Ooguri:2009cv}\cite{Donos:2009zf}\cite{Jeong:2009aa} or without supersymmetries \cite{Herzog:2008wg}\cite{Maldacena:2008wh}\cite{Adams:2008wt}. The simplest one is to consider the DLCQ of the AdS space. The compactification of the light-cone direction renders the isometry of the AdS space down to the Schr\"odinger group. In that case, as we have mentioned in the previous section, our deformations can also be applied in the DLCQ of the AdS space. We, however, prefer the geometry with $\frac{dt^2}{z^4}$ term as in \eqref{schr}. 

We claim that our deformations are universal in the sense that they can be applied to any (known) solutions with the Schr\"odinger invariance embedded in the supergravity. In this sense, our exact time-deformations are universal and the dual field theory interpretations should be valid in all the field theory duals of such geometries. In other words, the existence of such exact time-dependent deformations are common features of the dual field theories.

We begin with the deformation given by $c_1(t)$. As we observed, the deformation by $c_1(t)$ is actually locally trivial (up to a possible change of the boundary condition). It is simply induced by the coordinate transformation of the geometry $\xi \to \xi + \frac{1}{2}\int^t ds c_1(s)$. Since this is just a coordinate transformation, there is no obstacle to do it in any supergravity embeddings. Obviously we can do the same thing in the DLCQ of the AdS compactification.

Now let us consider the non-trivial deformation given by $c_2(t)$. In the AdS compactifications of the supergravity, such a deformation, a generalization of the Kaigorodov space, was studied in \cite{Brecher:2000pa}\cite{Chu:2006pa}, where the deformation by $c_2(t)$ does not change the other equations of motion except for the $(tt)$ component of the Einstein equations which are solved by assigning a definite power of $z$. Similarly, in our case, it can be shown that the deformation does not change the equations of motion except for the $(tt)$ component of the Einstein equation in any supergravity compactification. Again, the $(tt)$ component of the Einstein equation is solved by assigning a definite power of $z$. Thus, the generalized time-dependent Kaigorodov deformation by $c_2(t)$ gives the exact deformations of the any known supergravity solutions with the Sch\"odinger invariance. We also note that the deformation by $c_2(t)$ does not change the equations motion for flux with the ansatz that is compatible with the Sch\"odinger invariance. We will explicitly see the structure below when we discuss a concrete example.

The deformations by $c_1(t)$ and $c_2(t)$ only deal with the metric. Now, we consider the deformation given by the vector field $\epsilon(t)$. In the typical supergravity embedding, the vector field in the effective $d+2$ dimensional compactifications comes from the flux in the supergravity action. The effective mass term comes from the non-trivial eigenvalue for the internal Laplacian for the flux. The details will vary with respect to the origin of the vector field, but quite generally we can find the solutions of the supergravity equations motion corresponding to the deformation $\epsilon(t)$ in the following manner.

We first investigate the equation of motion for the flux. Thanks to the Schr\"odinger invariance and our ansatz for the time-dependent flux, $g_{tt}$ component of the metric does not affect the flux equation motion at all. Furthermore, the time-dependence in the flux does not affect the equations of motion, so we regard the time-dependence of the flux as if it were constant. This agrees with our feasibility to introduce an arbitrary function $\epsilon(t)$ in the vector field.  The flux equation motion, as a consequence, determines the $z$ dependence of the flux ansatz as in our effective field theory approach in section 2. 

The introduction of the time-dependence in the flux (as $\epsilon(t)$) will back-react to the geometry in the Einstein equation. We note that the backreaction only affects the $(tt)$ component of the Einstein equation. In addition, the $(tt)$ component of the Einstein equation can be solved by introducing the $(tt)$ component of the metric deformations of order $\epsilon^2(t)$. A crucial point here is that this $(tt)$ component of the metric, which is PP-wave type, will not affect all the other components of Einstein equations as well as the flux equations of motion, so by a suitable choice of the $g_{tt}$ deformation, we are able to solve all the equations motion induced by the flux deformation induced by $\epsilon(t)$. 

\subsection{Concrete example}

To make the above argument concrete, let us consider a particular M-theory compactification of the Schr\"odinger space $(d=1)$ based on the Sasaki-Einstein 7-fold. The original solution \cite{Donos:2009xc} is given by
\begin{align}
ds^2 &= \frac{1}{4}\left(q \frac{dt^2}{z^4} - \frac{2 dt d\xi}{z^2} + \frac{dx^2+dz^2}{z^2}\right) + ds^2(SE_7) \cr
G &= -\frac{3}{2^3z^4} dt \wedge d\xi \wedge dx \wedge dz + dt \wedge d\left(\frac{\tau}{2z^2}\right) \ , \label{gu}
\end{align}
where $\tau$ is a two-form on the Sasaki-Einstein 7-fold.
The flux equations of motion demand
\begin{align}
-*d* d\tau &= 24\tau \ ,  \ \ d*\tau = 0 \label{fleq} \ ,
\end{align}
where the Hodge star $*$ act on the internal space $SE_7$.
The $(tt)$ component of the Einstein equation demands
\begin{align}
D^i D_i q + 40q = -16|\tau|^2 - |d\tau|^2 \ ,
\end{align}
where $D_i$ is a covariant derivative on the internal space $SE_7$.
All the other equations motions are solved by \eqref{gu}. 
It is clear that the first terms in $G$ gives the effective cosmological constant and the second term gives the effective massive vector field whose mass squared is determined by the flux equation of motion \eqref{fleq}. 

In particular, for $SE_7=\mathbb{S}^7$, the explicit form of $\tau$ was constructed in \cite{Donos:2009xc} by splitting the parent $CY_4 = \mathbb{R}^8$ into $\mathbb{R}^8 = \mathbb{R}^4\times \mathbb{R}^4$ and considering a sum of terms which are $(1,1)$ with respect to the complex structure and primitive on one factor with a factor $dx^i$ on the other. In this simplest case, $q$ is a constant, which does not depend  on the internal coordinate, but in general it may depend on the internal coordinate.

As discussed above, we can introduce the $c_1(t)$ and $c_2(t)$ deformation by simply replacing the metric with
\begin{align}
ds^2 &= \frac{1}{4}\left[\left(q -  c_1(t) z^2 - c_2(t) z^5 \right)\frac{dt^2}{z^4} - \frac{2 dt d\xi}{z^2} + \frac{dx^2+dz^2}{z^2}\right] + ds^2(SE_7) \ . \label{eef}
\end{align}
The only equation of motion affected by the deformation is the $(tt)$ component of the Einstein equation, and it determines the power of $z$ appearing in the first term of \eqref{eef}. 
As discussed in the last subsection, we can explicitly see that the introduction of $c_1(t)$ and $c_2(t)$ does not change the other equations motion.

We now studies the flux deformation. 
\begin{align}
G &= -\frac{3}{2^3z^4} dt \wedge d\xi \wedge dx \wedge dz + dt \wedge d\left(E(t,z)\frac{\tau}{2 z^2}\right) \ . 
\end{align}
Since we identified the second term in $G$ as the effective massive vector field in the Schr\"odinger geometry, the introduction of $E(t,z)$ is natural.
The 11-dimensional flux equations motion 
\begin{align}
d*G + G\wedge G = 0
\end{align}
is insensitive to the $t$ dependence in $E(t,z)$. The $z$ dependence is uniquely fixed by
\begin{align}
E(t,z) = 1 + \epsilon(t)z^{5} \ .
\end{align}
together with \eqref{fleq}. Finally, the $(tt)$ component of the Einstein equation is now sourced by the new term coming from $\epsilon(t)$. This will be solved by introducing order $\epsilon^2(t)$ term in $g_{tt}$. The modification of $g_{tt}$ as well as the introduction of $\epsilon(t)$ in the flux do not affect all the other equations of motion so that we find the embedding of the solutions within the M-theory compactification.

It is clear that the above construction can be repeated in other string/M-theory realizations of the Schr\"odinger space-time. The only non-trivial part is to identify the effective massive vector field in the flux ansatz. 

\section{Relativistic case}
In section 2.3, we have studied the time-dependent scale invariant and/or conformally invariant deformations of the Schr\"odinger invariant field theory.
A similar question can be addressed in the relativistic AdS/CFT correspondence.  From the field theory viewpoint, it seems possible to deform the relativistic CFT so that only the translational invariance is broken while preserving dilation, special conformal transformation as well as Lorentz transformation. In the Lagrangian description, for instance, we may add the interaction
\begin{align}
 \int d^d x (x^2) O(x) \ ,
\end{align}
where $O(x)$ is a primary scalar operator of conformal dimension $d+2$. At the first order, such deformations are scale and conformal invariant as well as Lorentz invariant, but they break the translational invariance due to the explicit $x^2$ dependence in the interaction.

Are there corresponding deformations in the CFT side? Surprisingly it seems very difficult to introduce such deformations contrary to the naive expectation from the field theory discussions. We first note that the special conformal transformation acts as ($a=0,1,\cdots ,d$)
\begin{align}
\delta x_a = 2(\epsilon^bx_b) x_a - (z^2 + x^bx_b) \epsilon_a \ , \ \ \delta z = 2(\epsilon^bx_b)z \ ,
\end{align}
so while $x^2/z^2$ is invariant under the dilation, it is not invariant under the special conformal transformation. In particular, any scalar field profile like $\phi = f(x^2/z^2)$ is invariant under the scale transformation, but not invariant under the special conformal transformation. Indeed, as a simple corollary, there would be no such scalar deformations possible from the bulk theory viewpoint.

While there is no invariant scalar perturbation, there are possible vector or metric perturbations. Let us consider the vector field profile
\begin{align}
A = A_{\mu}dx^\mu =  K\left(\frac{x^2}{z^2}\right) \frac{dz}{z} + J\left(\frac{x^2}{z^2}\right) \frac{x_a dx^a}{z^2} \ . \label{vecd}
\end{align}
Again because of the non-invariance of $x^2/z^2$ under the special conformal transformation, each terms in \eqref{vecd} are not invariant under the special conformal transformation while it is obviously scale invariant.
In order for the deformation to be conformally invariant, $K(y)$ and $J(y)$ should satisfy
\begin{align}
K'(y) (1+y) + J(y) &=  0 \cr
2K(y) + J(y)(-1+y) &= 0 \cr
J'(y)(1+y) + J &= 0 \ .
\end{align}
This is an overdetermined system, but it has a solution:
\begin{align}
K(y) = \frac{c}{2} \frac{1-y}{1+y} \cr
J(y) = \frac{c}{1+y} \ ,
\end{align}
where $c$ is an integration constant.

Similarly, we can study the following metric perturbations:
\begin{align}
ds^2 = A\left(\frac{x^2}{z^2}\right) \frac{dz^2}{z^2} + B\left(\frac{x^2}{z^2}\right)\frac{dx^2}{z^2} + C\left(\frac{x^2}{z^2}\right) \frac{x_a dx^a dz}{z^3} + D\left(\frac{x^2}{z^2}\right) \frac{(x_a dx^a)^2}{z^4} \ . 
\end{align}
The invariance under the special conformal transformation demands
\begin{align}
(1+y)A'(y)  + C(y) &= 0 \cr
4A(y) - 4B(y) + (y-1)C(y) &= 0 \cr
(1+y)B'(y) &= 0 \cr
(1+y)C'(y) + C(y) + 2D(y) &= 0 \cr
C(y) + (-1+y) D(y) &= 0 \cr
(1+y)D'(y) + 2 D(y) & = 0 \ .
\end{align} 
Again, the equations look overdetermined, but there is a nontrivial solution:
\begin{align}
A'(y) &= c'\frac{y-1}{(1+y)^3} \cr
B'(y) &= 0 \cr
C(y)  &= c'\frac{1-y}{(1+y)^2} \cr
D(y)  & = c'\frac{1}{(1+y)^2} \ 
\end{align}
with another integration constant $c'$.

We note that we have not imposed any equations of motion yet. In this sense, the solutions are quite restrictive. It is clear that arbitrary actions and equations of motion do not allow such a solution, and in general, there would be no scale and conformal invariant deformations of the relativistic conformal field theory that break the translational invariance from the gravity viewpoint. Again, this may be related to the fact that the representation theory of the relativistic conformal algebra has a non-trivial involution anti-automorphism replacing the momentum $P^\mu$ with the special conformal generator $K^\mu$, so it is likely that the difficulty to find a scale and conformal invariant but non-translational invariant deformations may be related to the difficulty to find a scale invariant but non-conformal deformations of the relativistic conformal field theories.

\section{Discussions and Summary}
What happens if we hit the singularity of the universe? Is it the end of the universe, or will the stringy effects remove it? How does the time begin or end? These are fundamental questions that should be addressed and hopefully answered in fundamental theories of quantum gravity. If the holography is one of the most fundamental nature of the gravity, the dual field theory approach would enable us to answer the question.

The AdS/CMP correspondence might give us a novel way to probe the singularities of the universe from our lab experiments. Our time-dependent deformations of the non-relativistic AdS/CFT correspondence contain the PP-wave singularity. Since they are universal, one may expect that they are realized in, for example, the cold atoms, or unitary fermion system. 

In this paper, we have also clarified that the distinction between the scale invariance and the conformal invariance is more manifest in the time-dependent non-relativistic system. Although in the unitary Poincar\'e invariant field theories, or in the unitary Galilean invariant field theories, these two concepts might be equivalent, we have found otherwise in the time-dependent background from the gravity approach. It would be interesting to confirm our result from the field theory approach.

The subtle relation between the conformal invariance and the scale invariance is a fundamental question in theoretical physics that is yet to be solved. The probelm is one of the few good examples that can be shared by string theorists and the condensed matter physicists. We hope that further collaborations will give a theoretical as well as experimental clue to this elementary question that lies in the basic foundation of the world sheet formulation of the string theory, quantum gravity, as well as critical phenomena in condensed matter physics.

\section*{Acknowledgements}
The author would like to thank Gabriel Wong for the collaboration in the early stage of this work. He also thanks Ben Freivogel for the discussion of the PP-wave singularity.
The work was supported in part by the National Science Foundation under Grant No.\ PHY05-55662 and the UC Berkeley Center for Theoretical Physics  and World Premier International Research Center Initiative (WPI Initiative), MEXT, Japan. 

\appendix

\section{non-relativistic conformal algebra}

We summarize the non-relativistic conformal algebra \cite{Hagen:1972pd}\cite{Hussin:1986cc} in (1+2) dimension. The higher dimensional analogue will be obvious:
\begin{align}
& i[J,P_+] = -i P_+ \ , \ \ \ \ \ i[J,P_-] = + i P_- \ , \ \ \ \ \ \ i[J,G_+] = - iG_+ \ , \ \ \ \ i[ J ,G_-] = + iG_- \ ,\cr
& i[H,G_+] = + P_+ \ ,  \ \, \ \ \ i[H,G_-] = + P_- \ , \ \ \ \ i[K,P_+] = - G_+ \ ,  \ \ \ \ \ i[ K,P_-] = - G_- \ ,\cr
& i[D,P_+] = - P_+ \ , \ \ \ \ \ i[D,P_-] = - P_- \ , \ \ \ \ \ i[D,G_+] = + G_+ \ ,  \ \ \ \ \ i[ D,G_-] = + G_- \ ,\cr
& i[H, D] = \ 2H \ , \ \ \ \ \ \ \ \ i[H,K] =  \ D \ , \ \ \ \ \ \ \ \ \ i[D,K] = \ 2K \ ,  \ \ \ \ \ \ \ \ \ \ i[P_+,G_-] = \ 2M \ .
\end{align}
In our notation, $H$ is the non-relativistic Hamiltonian, $P_\pm = P_x \pm i P_y$ are the momentum, $J$ is the U(1) angular momentum, $D$ is the dilatation, $K$ is the special conformal transformation, and $G_{\pm} = G_x \pm iG_{y}$ are the Galilean boost generators. Moreover, $M$ is the total mass generator. The total mass $M$ is related to the particle number by a proportional factor: $M = mN = m\int d^d x \rho $.

We note that the non-relativistic superconformal algebra has a grading structure with respect to the dilatation operator $D$ and can be triangular-decomposed as
\begin{equation}
{\cal A}_+ \oplus {\cal A}_0 \oplus {\cal A}_-,
\end{equation}
where
\begin{eqnarray}
&& {\cal A}_+ = \{ \ P_- , \ {P}_+, \ H  \ \} \nonumber \\
&& {\cal A}_0 = \{ \ J, \ M, \ D \ \} \nonumber \\
&& {\cal A}_- = \{ \ G_-, \ {G}_+, \ K \ \}.
\end{eqnarray}
We also notice that the non-relativistic conformal algebra has a non-trivial involution anti-automorphism of the algebra \cite{Nakayama:2008qm} given by
\begin{eqnarray}
&& w(J)= J, \qquad \ \ \ w(P_\pm)= {G}_\mp, \qquad \ w(G_\pm)={P}_\mp, \qquad  w(H) =-K, \nonumber \\
&& w(R) = R, \qquad \ w(D) = -D, \qquad \ w(M) = -M, \qquad \ w(K) = -H \ .
\end{eqnarray}
This anti-automorphism is essential in the ``radial" quantization of non-relativistic conformal field theories \cite{Nishida:2007pj}.

\end{document}